\documentclass[conference]{IEEEtran}
\IEEEoverridecommandlockouts
\usepackage[]{mdframed}
\usepackage{cite}
\usepackage{amsmath,amssymb,amsfonts}
\usepackage{algorithmic}
\usepackage{graphicx}
\usepackage{textcomp}
\usepackage{xcolor}
\usepackage{longtable}
\usepackage{hyperref}

\def\BibTeX{{\rm B\kern-.05em{\sc i\kern-.025em b}\kern-.08em
    T\kern-.1667em\lower.7ex\hbox{E}\kern-.125emX}}
\begin{document}

\title{Currency Arbitrage Optimization using Quantum Annealing, QAOA and Constraint Mapping  \\

}

\author{
\IEEEauthorblockN{Sangram Deshpande \\and Elin Ranjan Das}
\IEEEauthorblockA{\textit{Department of Electrical and Computer Engineering} \\
\textit{North Carolina State University}\\
\textit{Raleigh, North Carolina, 27695-8206} \\
\textit{Email: ssdesh24, edas@ncsu.edu
}}

\and

\IEEEauthorblockN{Frank Mueller}
\IEEEauthorblockA{\textit{Department of Computer Science}\\
\textit{North Carolina State University}\\
\textit{Raleigh, North Carolina, 27695-8206}\\
\textit{Email: fmuelle@ncsu.edu}}
}

\maketitle

\begin{abstract}
  Currency arbitrage capitalizes on price discrepancies in currency
  exchange rates between markets to produce profits with minimal risk.
  By employing a combinatorial optimization problem, one can ascertain
  optimal paths within directed graphs, thereby facilitating the
  efficient identification of profitable trading routes. This research
  investigates the methodologies of quantum annealing and gate-based
  quantum computing in relation to the currency arbitrage problem. In
  this study, we implement the Quantum Approximate Optimization
  Algorithm (QAOA) utilizing Qiskit version 1.2. In order to optimize
  the parameters of QAOA, we perform simulations utilizing the
  AerSimulator and carry out experiments in simulation.
  Furthermore, we present an NchooseK-based methodology utilizing
  D-Wave's Ocean suite.  This methodology enables a comparison of the
  effectiveness of quantum techniques in identifying optimal arbitrage
  paths. The results of our study enhance the existing literature on
  the application of quantum computing in financial optimization
  challenges, emphasizing both the prospective benefits and the
  present limitations of these developing technologies in real-world
  scenarios.

\end{abstract}

\begin{IEEEkeywords}
arbitrage, annealing, profit, NchooseK, QAOA, optimization
\end{IEEEkeywords}

\section{Introduction}
Currency arbitrage is a trading strategy that takes advantage of price
discrepancies for the same currency pair across different markets or
exchanges. Market participants have the opportunity to acquire a
currency at one rate in one marketplace and subsequently sell it at
another one. If the former price is lower than the latter, they can
realize a profit from the price differential, otherwise a loss. A
strategy to make profits can take on multiple manifestations,
including triangular arbitrage, wherein traders capitalize on
discrepancies among three currencies, or through direct transactions
involving two currencies. The primary objective is to leverage market
inefficiencies in order to achieve profits while maintaining a low
level of risk.

The classical Bellman-Ford algorithm serves as a method to detect
arbitrage opportunities. This algorithm facilitates the identification
of negative cycles within a graph that models currency exchange
rates. It is crucial to determine the presence of negative cycles
within a graph and, if they exist, to identify the particular segments
that contribute to these anomalies~\cite{b1}.

The identification of currency arbitrage opportunities can be
conceptualized as an optimal path search within a directed graph,
wherein nodes symbolize currencies and edge weights reflect exchange
rates. This introduces a combinatorial optimization challenge that
necessitates the development of efficient algorithms to facilitate
prompt decision-making.

\begin{figure*}
    \centering
    \includegraphics[width= \textwidth]{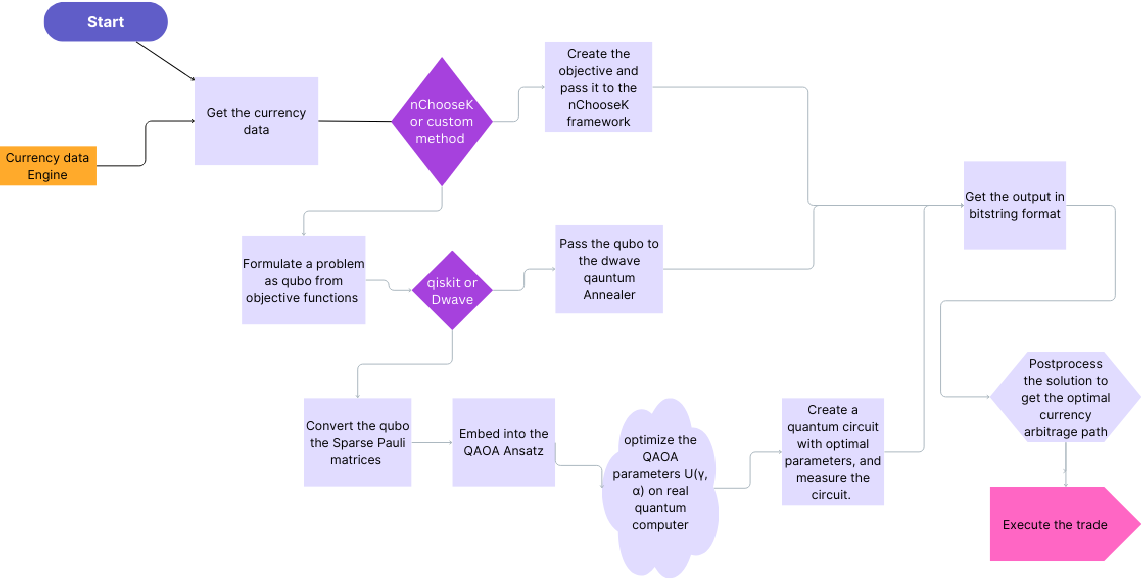}
    \caption{Flowchart of the Algorithm}
    \label{fig:f1}
\end{figure*}

An algorithm following this concept is given in the flowchart in
Figure~\ref{fig:f1}. We expressed the combinatorial optimization
problem as a quadratic unconstrained binary optimization (QUBO)
problem for resolution on the D-Wave quantum annealer, and we devised
a formulation for QAOA to implement it on IBM quantum simulator or
actual device hardware.  We additionally reshaped the problem into a
combinatorial problem suitable for a domain-specific language,
NchooseK, and exposed its solution to quantum annealers and
gate-based quantum computers.  We conducted a comparative study of
the circuit depth associated with NchooseK and that of QAOA for the
identical problem, demonstrating that NchooseK provides a notable
enhancement over current QAOA frameworks.

\section{Related Works}

Currency arbitrage has been a subject of interest in the context of
quantum computing. Notably, several quantum-inspired algorithms have
been applied to identify optimal arbitrage paths.  One such approach
utilizes the Simulated Bifurcation (SB) algorithm, a quantum-inspired
method designed to solve combinatorial optimization problems
efficiently. Tatsumura et al.~\cite{b2} have developed a currency
arbitrage machine based on the SB algorithm capable of rapidly
detecting optimal exchange paths among multiple currencies. This
system demonstrates the potential of SB in financial applications,
particularly in capturing short-lived arbitrage opportunities.

Another significant contribution comes from Carrascal et al.~\cite{b3},
who applied the Variational Quantum Eigensolver (VQE) algorithm to the
currency arbitrage problem. The authors implemented a Differential
Evolution (DE) optimizer as a substitute for the conventional COBYLA
solver. Their findings demonstrate that the DE-based method
successfully converges to the optimal solution in contexts where other
frequently employed optimizers, such as COBYLA, encounter difficulties
in locating the global minimum. This indicates that the combination of
evolutionary algorithms and quantum optimization techniques may
improve convergence characteristics in intricate financial challenges.
The findings highlight the capabilities of quantum and
quantum-inspired algorithms in tackling intricate financial
optimization issues, presenting valuable opportunities for further
investigation and real-world implementations in currency arbitrage.

\section{Design}

\subsection{QUBO Formulation}

We formulated QUBOs utilizing Python. The arbitrage problem was
reshaped as a problem in graph theory, wherein a directed graph is
constructed such that each node signifies a currency, and each
directed edge is assigned a weight corresponding to the relevant
conversion rate. Our objective is to identify a cyclic path within an
asset and exchange rate graph that yields the maximum profit rate. The
profit rate is defined as the product of exchange rates from the
$i^{th}$ currency to the $j^{th}$ currency, denoted as $(r_{ij})$. A
decision variable, $b_{ij}$, is defined such that it takes the value
of 1 if the corresponding edge $(i, j)$ is included in the selected
cycle and 0 if not. The profit rate is denoted as
$\Pi r^{b_{ij}} _{ij}$. This represents a polynomial whose degree
corresponds to the quantity of conversion rates, also referred to as
edges. By optimizing the logarithm of the product, the order can be
reduced to linear. We proceed to introduce a linear cost
function~\cite{b4},
\begin{equation}
    C = \sum_{i,j}-log(r_{ij})b_{ij},
\end{equation}
and a penalty function for cyclic constraints \cite{b1},
\begin{equation}
\begin{aligned}
     P = \sum_{i}\sum_{j\neq j'} b_{i,j}b_{i,j'} +
    \sum_{j}\sum_{i\neq i'} b_{i,j}b_{i',j}.
\end{aligned}
\end{equation}
The first and second terms correspond to one incoming and one outgoing
edge per currency, respectively, to ensure a closed arbitrage cycle.
The total cost function, $C_{tot}$, is a linear combination of C and
P, namely
\begin{equation}
    C_{tot} = C + m_pP,
\end{equation}
where $m_p$ is a constraint factor. Various values of $m_p$ were
employed, and an appropriate value was selected for inclusion in the
objective function.

\begin{table}[h!]
\caption{Four Currency Exchange Rates}
\label{table:rate4}
\centering
\renewcommand{\arraystretch}{1.5}
\setlength{\tabcolsep}{10pt}
\begin{tabular}{|c|c|c|c|c|}
\hline
 & EUR & USD & CHF & JPY \\ \hline
EUR & 1.0 & 1.13217 & 1.11777 & 120.756 \\ \hline
USD & 1/1.13403 & 1.0 & 0.98804 & 106.034 \\ \hline
CHF & 1/1.12005 & 1/0.99250 & 1.0 & 105.564 \\ \hline
JPY & 1/120.887 & 1/106.266 & 1/108.042 & 1.0 \\ \hline
\end{tabular}
\end{table}

A sample currency exchange rate table from~\cite{b2} is utilized,
encompassing four distinct currencies (see
Table~\ref{table:rate4}). The information was subsequently encoded
into a QUBO dictionary, which was further refined to incorporate the
penalty function. We additionally utilize tables for five and six
currency exchange rates in Tables~\ref{table:rate5}
and~\ref{table:rate6}, respectively.

\begin{table}[h!]
\caption{Five Currency Exchange Rates}
\label{table:rate5}
\centering
\renewcommand{\arraystretch}{1.5}
\setlength{\tabcolsep}{10pt}
\begin{tabular}{|c|c|c|c|c|c|}
\hline
 & USD & EUR & GBP & JPY & AUD \\ \hline
USD & 1.0 & 0.8953 & 0.7682 & 148.76 & 1.5213 \\ \hline
EUR & 1.1170 & 1.0 & 0.8586 & 166.06 & 1.6993 \\ \hline
GBP & 1.3015 & 1.1645 & 1.0 & 193.40 & 1.9801 \\ \hline
JPY & 0.0067 & 0.0060 & 0.0052 & 1.0 & 0.0102 \\ \hline
AUD & 0.6572 & 0.5885 & 0.5050 & 97.88 & 1.0 \\ \hline
\end{tabular}
\end{table}

\begin{table}[h!]
\caption{Six-Currency exchange rate table}
\label{table:rate6}
\centering
\renewcommand{\arraystretch}{1.5}
\setlength{\tabcolsep}{7.5pt}
\begin{tabular}{|c|c|c|c|c|c|c|}
\hline
 & USD & EUR & GBP & JPY & AUD & CAD \\ \hline
USD & 1.0 & 0.8953 & 0.7682 & 148.76 & 1.5213 & 1.3407 \\ \hline
EUR & 1.1170 & 1.0 & 0.8586 & 166.06 & 1.6993 & 1.4971 \\ \hline
GBP & 1.3015 & 1.1645 & 1.0 & 193.40 & 1.9801 & 1.7433 \\ \hline
JPY & 0.0067 & 0.0060 & 0.0052 & 1.0 & 0.0102 & 0.0090 \\ \hline
AUD & 0.6572 & 0.5885 & 0.5050 & 97.88 & 1.0 & 0.8814 \\ \hline
CAD & 0.7457 & 0.6684 & 0.5738 & 111.23 & 1.1345 & 1.0 \\ \hline
\end{tabular}
\end{table}

\subsection{Translating QUBO to NchooseK}

NchooseK~\cite{b5} is a constraint-based programming model and a
specific type of integer linear programming (ILP). Programs are
composed of Boolean variables along with a collection of constraints
imposed on them. There are two types of constraints, hard constraints
and soft constraints. According to the definitions in~\cite{b6}, ``An
NchooseK hard constraint, written as \textit{nck(N,K),} consists of a
variable collection \textit{N} and a selection set \textit{K}. It is
satisfied if the cardinality of the variable collection whose
variables are TRUE equals one of the numbers in the selection set
\begin{equation}
    nck(N,K) = (\sum_{n\in N}n)\in K,
\end{equation}
where $n \in \{0,1\}$ and we associate FALSE with 0 and TRUE with 1.
An NchooseK soft constraint, written as \textit{nck(N,K,soft)}, acts
as a desired but not required constraint.''

Consider Table~\ref{table:rate4} again. We observe that to obtain a
closed arbitrage path, exactly one edge $(r_{i,j})$ from each row and
column is assigned a \textit{true} decision variable $b_{i,j} =
1$. This is implemented as a hard constraint, ensuring that each
currency node has precisely one outgoing edge and one incoming
edge. However, this formulation only identifies all possible arbitrage
cycles without prioritizing the most profitable one. To address this,
soft constraints are introduced to bias the solution towards
maximizing profit. The logarithmic value of each currency pair's
exchange rate is mapped to the soft constraint weight associated with
its decision variable. These soft constraints are added to encourage
the selection of edges corresponding to higher profit rates. The
scaling factor, applied to the logarithmic profit bias, determines the
number of times the soft constraints are reinforced. In the
implementation, the algorithm iteratively composes soft constraints by
repeating them based on the calculated profit bias. This process
effectively integrates the soft constraints with the hard constraints
to form the overall optimization formulation. The higher the profit
bias of a currency pair, the stronger the influence of its
corresponding soft constraint, increasing the likelihood of including
the edge in the final solution. The constrained problem is
subsequently solved using the \textit{ocean.solver()} function, which
utilizes the D-Wave Ocean Solver (D-Wave Advantage V4.1) for quantum
annealing. This process identifies the most profitable arbitrage cycle
under the given constraints.

\subsection{Translating QUBO to IBM Quantum native format}

With the QUBO generated, we can translate it into the IBM Quantum
Qiskit native representation. This translation is crucial as it
ensures compatibility with IBM's quantum computing framework. The
resulting native QUBO will then serve as input for the QAOA Ansatz.

Employing the QAOA Ansatz, a QAOA quantum circuit will be developed,
aimed at effectively navigating the solution space delineated by the
QUBO. It is possible to develop a custom QAOA Ansatz capable of
accommodating a substantial number of variables, which is expected to
yield a quantum circuit. The circuit will subsequently be executed on
IBM Quantum devices, which are superconducting gate-based systems
specifically designed for the execution of quantum algorithms such as
QAOA.

After execution, the output will be obtained in the form of binary
bitstrings, representing the optimal or near-optimal solutions to the
original problem. These results can then be analyzed and interpreted
after the fact, allowing for insights into the best configurations of
the selected variables. Overall, this process illustrates the seamless
integration of classical optimization methods with cutting-edge
quantum technology paving the way for solving complex problems that
are otherwise intractable.

Next, we define a cost function that reflects the profitability of
completing a cycle, which transforms multiplicative relationships into
an additive format for easier optimization. We evaluate the efficiency
of QAOA in solving this problem by conducting a comparative study with
quantum annealing solutions. This study aims to assess the performance
of both quantum formulations in identifying the optimal cycle by
analyzing factors such as computation time, accuracy, and scalability
to determine which approach yields superior results in the context of
currency arbitrage.

\begin{figure*}
    \centering
    \includegraphics[width=\textwidth]{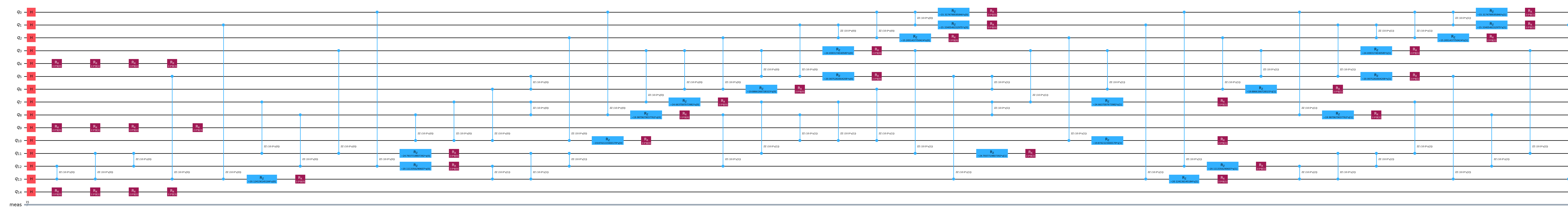}
    \caption{Quantum Circuit with Optimal Parameters, with 1 Layer }
    \label{fig:circuit_diagram}
\end{figure*}

\section{Implementation}

We began by formulating the QUBO problem to maximize arbitrage
profits. Utilizing D-Wave's Leap service, we submitted the QUBO to the
quantum annealer by conducting experiments with 1,000 shots (repeated
experiments with a measurement outcome) per run to ensure statistical
significance.

For gate-based quantum computing, we employed QAOA, recognized for its
efficiency in combinatorial optimization. Using Qiskit, we executed
the algorithm on IBM's quantum simulator (or actual device hardware),
determining optimal parameters through iterative refinement and
performing measurements in the Z-basis to extract meaningful results.

The NchooseK algorithms were developed using Python's NchooseK
domain-specific language (DSL), facilitating the expression of
combinatorial constraints inherent in the currency arbitrage
problem. To ensure compatibility with our problem requirements, we
adapted our implementation to an earlier version of Qiskit, aligning
with the specific functionalities needed for our approach.


Upon completing all experiments, the outputs (represented as
bit-strings) were post-processed to identify the optimal currency
arbitrage paths that maximize profit. This post-processing involved
interpreting the bit-strings to determine specific sequences of
currency exchanges corresponding to profitable cycles. By analyzing
these sequences, we could pinpoint arbitrage opportunities that
exploit discrepancies in exchange rates across different markets. This
approach ensures that the derived trading strategies are both
actionable and aligned with the goal of achieving maximum profit
through arbitrage.

\section{Experimental Setup}

We tried to solve this problem with D-Wave Quantum Annealer, NchooseK
solver, and IBM Quantum's Qiskit package. The flowchart demonstrating
this methodology is given in Figure~\ref{fig:f1}.

Utilizing the \textit{dimod} package, we created a QUBO objective
function that incorporated the cost function and the penalty function
for the constraints. We then solved it using the D-Wave
\textit{EmbeddingComposite} sampler with 1,000 shots per run. The
sampler subsequently returned the solution with the lowest
energy. Next, we extract the currency exchange path in a closed-cycle
format. If the calculated profit exceeds 1, the arbitrage path is
printed; otherwise, the output indicates that there is no optimal
arbitrage path.

In the NchooseK Ocean Solver, we employed 1,000 shots and allocated 200
microseconds of annealing time. This enables the algorithm to navigate
the landscape more comprehensively. The soft constraint weight scaling
factor was established at 100. This experiment was repeated 6 times to
ensure accuracy of the results. The best results are discussed in the
results section.

In our implementation of QAOA, we utilized 1,024 shots and employed
the AerSimulator for running the quantum circuit. We also aimed to
execute the QAOA circuit on real quantum hardware. However, when
running on the AerSimulator in noiseless settings, we observed
suboptimal results. For classical optimization of parameters within
the QAOA circuit, we employed the COBYLA optimizer of the SciPy
framework. While Quantum Gradient Descent could have been an
alternative, it would have significantly increased the circuit depth.

The COBYLA optimizer aids in the convergence of parameters within QAOA
by iteratively refining them over repeated subcircuits. In this case,
the optimizer required 92 iterations to converge and determine the
optimal parameters for the given objective and constraints. While the
convergence was relatively slow during the initial iterations, the
process ultimately yielded suboptimal results after 100
iterations. The objective was to minimize the cost function, and the
optimized value achieved was -29.242109307796706.




A simplified QAOA circuit with just 1 layer is shown in
Figure~\ref{fig:circuit_diagram}, whereas the actual circuit consists
of 4 such layers, significantly increasing its depth, size, and
width. In comparison, 100 iterations would correspond to a variational
optimization loop where the QAOA parameters (2 angles per layer) are
updated iteratively to improve the solution quality. The number of
circuit layers (4 in this case) determines the circuit depth in a
single iteration, while the 100 iterations refer to the classical optimization
steps applied to the parameters across multiple runs of the
circuit. Thus, the two metrics --- layers and iterations --- are distinct but
interdependent, as deeper circuits typically require more iterations
to converge to an optimal solution.

We determined the characteristics of the QAOA circuit after mapping it
to simulators and refining its parameters. The finalized configuration
is as follows: The circuit has a depth of 44 parallel gates along the
critical path for a total of 234 gates over 30 qubits. These
parameters, while reflecting the circuit's structure, also pose
significant challenges.

\begin{table*}[tb]
\caption{Currency Arbitrage Experimental Results Across Different Quantum Computing Frameworks}
\centering
\begin{tabular}{c|c|c|c|c}
\hline  

\textbf{Sr. No.} & \textbf{Currencies} & \textbf{Experiment mode} & \textbf{Currency Arbitrage Path} & \textbf{Profit Rate} \\ \hline  

1.  & 4 & D-Wave QUBO Solver & EUR $\to$ JPY $\to$ USD $\to$ CHF $\to$ EUR & 1.002424106050562 \\ \hline

2.  & 4 & NchooseK Ocean Solver & EUR $\to$ JPY $\to$ USD $\to$ CHF $\to$ EUR & 1.002424106050562 \\ \hline
3.  & 4 & NchooseK qiskit Solver & GBP$\to$AUD $\to$ CAD $\to$INR $\to$GBP
& 0.9976094869628244 \\ \hline
4.  & 4 & Qiskit QAOA ran on IBM Q simulator  & EUR$\to$USD $\to$ EUR 
& 0.9983598317504827 \\ \hline

5.  & 5 & D-Wave QUBO Solver & USD $\to$ EUR $\to$ AUD $\to$ JPY $\to$ GBP $\to$ USD & 1.002309461549003 \\ \hline
6.  & 5 & NchooseK Ocean Solver & USD$\to$EUR $\to$ JPY $\to$GBP $\to$AUD $\to$USD
& 1.0005744453979277 \\ \hline
7.  & 5 & NchooseK Qiskit Solver & USD $\to$JPY $\to$ USD
& 1.009 \\ \hline

8.  & 6 & D-Wave QUBO Solver & USD $\to$ EUR $\to$ AUD $\to$ CAD $\to$ JPY $\to$ GBP $\to$ USD & 1.0039286603299495 \\ \hline
9.  & 6 & NchooseK Ocean Solver & USD$\to$JPY $\to$ GBP $\to$AUD $\to$USD
& 1.0011564702289195 \\ \hline

\end{tabular}
\label{tab:experiment_results}
\end{table*}

The width of 30 qubits can strain the capacity of current quantum hardware. Ensuring high-fidelity operations on all these qubits is critical, but noise and limited connectivity in hardware can degrade solution quality. Similarly, the depth of 44 highlights the sequential constraints on operations, which increases the likelihood of decoherence and cumulative gate errors during execution. The total number of 234 gates, roughly half of which are two-qubit gates with much higher noise influx than single-qubit gates, adds to the computational complexity and further amplifies noise susceptibility.

These factors collectively hinder the quality of the solution by introducing errors that propagate throughout the circuit. As depth and size increase, the reliability of results decreases, especially on near-term quantum devices with limited error correction. Balancing these metrics is crucial to achieving a feasible and accurate implementation. Exploring techniques such as circuit optimization, noise mitigation, or targeting hardware with advanced error correction could help improve the solution quality despite these constraints.

We determined the optimal parameters for each pair of $\gamma$ and
$\beta$ for a 4-layer QAOA circuit as:
\[
\begin{array}{|c|c|}
\hline
\gamma_1 = 1.569 & \beta_1 = 1.522 \\ \hline
\gamma_2 = 2.811 & \beta_2 = 2.556 \\ \hline
\gamma_3 = 1.555 & \beta_3 = 2.696 \\ \hline
\gamma_4 = 1.544 & \beta_4 = 1.591 \\ \hline
\end{array}
\]

These parameters were derived from the optimization techniques
designed to minimize the cost function of the problem. Once the
optimal solution was obtained, it was mapped to the currency pairs to
identify the most efficient arbitrage path.


\section{Results}

\begin{figure}[htbp]
    \centering
    \includegraphics[width=1\linewidth]{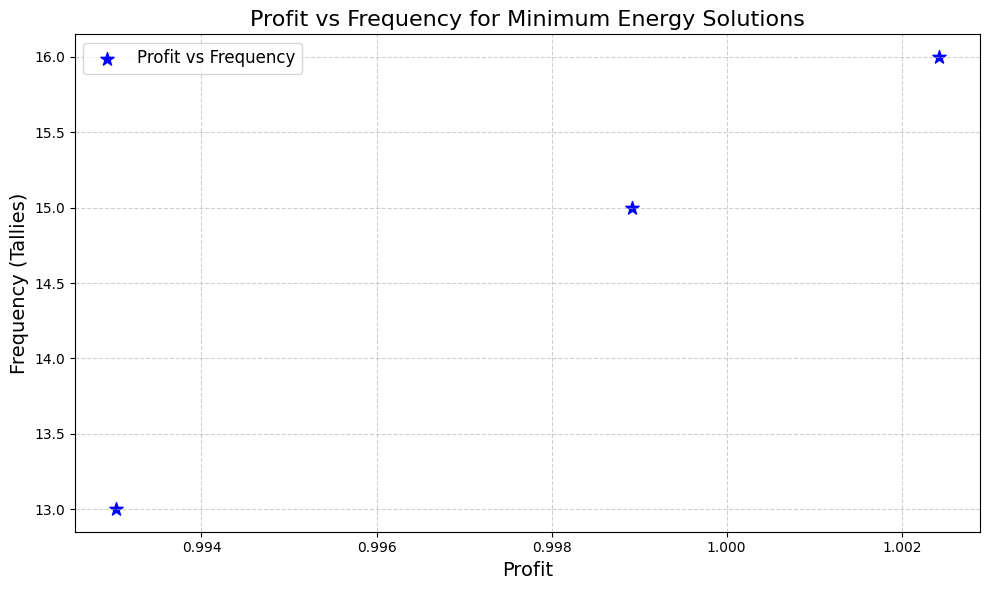}
    \caption{4-Currencies' Frequency of Annealing Shots vs. Profit of Degenerate Energy Solutions in NchooseK Ocean Solver. Notice that the Best Solution is included in the Solution Space.}
   \label{fig:f4}
\end{figure}

\begin{figure}[htbp]
    \centering
    \includegraphics[width=1\linewidth]{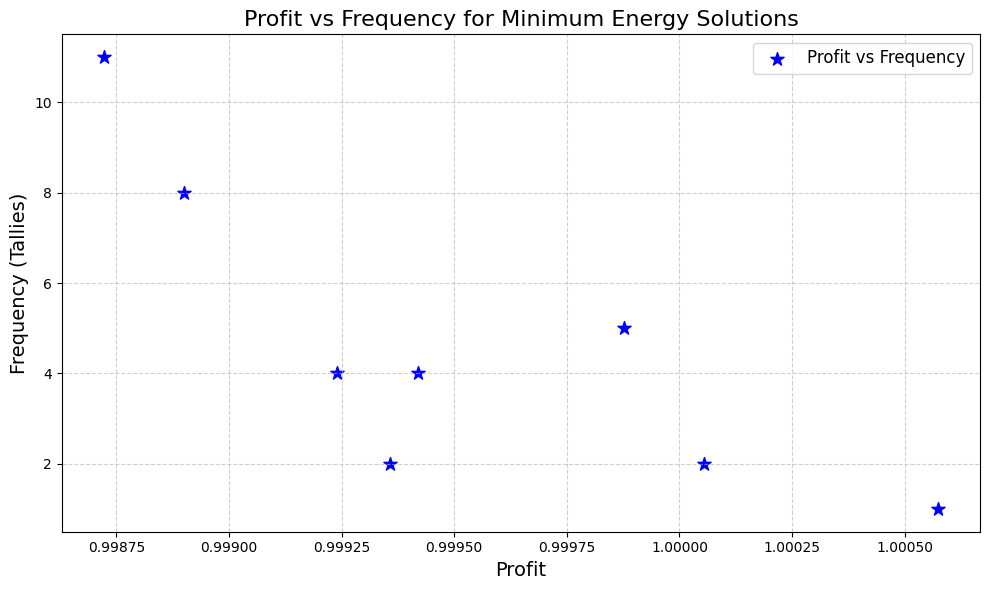}
    \caption{5-Currencies' Frequency of Annealing Shots vs. Profit of Degenerate Energy Solutions in NchooseK Ocean Solver.}
       \label{fig:f5}
\end{figure}

\begin{figure}[htbp]
    \centering
    \includegraphics[width=1\linewidth]{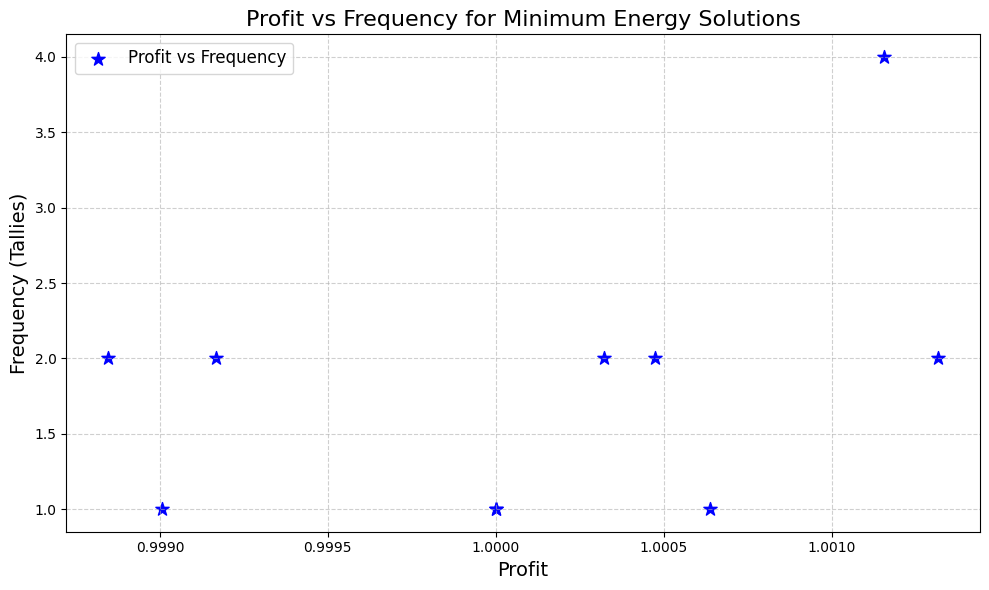}
    \caption{6 Currencies' Frequency of Annealing Shots vs. Profit of Degenerate Energy Solutions in NchooseK Ocean Solver.}
       \label{fig:f6}
\end{figure}

Table~\ref{tab:experiment_results} depicts the results of the D-Wave
QUBO Solver and the NchooseK Ocean Solver, both of which use the same
quantum annealing device.  Upon executing the problem using NchooseK,
the program returned degenerate solutions that exhibited identical
lowest energy levels. The frequency of these results was assessed, and
the profit for each solution was calculated as depicted in
Figures~\ref{fig:f4},~\ref{fig:f5}, and~\ref{fig:f6}. The best
solution is indeed included within the degenerate solution space.  For
the four currencies case, the minimum energy level was associated with
44 shots (out of 1,000), of which 16 represent the optimal cycle. When
we increased the number of currencies (five and six), the performance
deteriorated with increasing degeneracy. For five currencies, merely 3
out of the 37 shots at the lowest energy levels reveal the profitable
paths, with only 1 aligning with the optimal path. For six currencies,
11 out of 17 shots at the lowest energy levels yield profitability,
with 2 identified as optimal.  Although an increase in currency yields
fewer shots for the optimal path in the degenerate energy shots, the
fraction of profitable paths increases, thereby presenting the
opportunity to exploit various profitable paths as the combinatorial
problem increases exponentially in complexity. The frequency of the
degenerate minimum energy solutions varied in each experiment
iteration.

We also obtained the experimental results using Qiskit version 1.2.0,
but the ``solutions'' did not satisfy the specified constraints. The
best solution provided by QAOA on Qiskit for four currencies suggested
trading between only two currency pairs, EUR and USD, with the profit
= 0.9983598317504827, which is suboptimal, i.e., a net loss. This
limitation might be addressed by running the algorithm on a real
quantum processing unit (QPU) with the implementation of error
correction mechanisms. After all, if results where suboptimal under
simulation, we cannot expect them to improve on current
superconducting devices considering their level of noise.

\section{Future Works}

In this study, we illustrate that the formulation of a QUBO and its
resolution using D-Wave Quantum Annealers yielded the most efficient
solution in the least amount of time. Nevertheless, this methodology
presents certain difficulties as the QUBO formulation is dependent on
a sophisticated mathematical framework that lacks inherent
intuitiveness. Furthermore, the efficacy of the solution is
significantly affected by variables including the quantity of shots
and the prioritization of penalty constraints.

Similarly, the application of the QAOA algorithm to tackle this issue
revealed challenges regarding the convergence of classical solvers
from the Scipy package.  This issue aligns with findings from
Carrascal et al.~\cite{b3} who adopted a genetic algorithm technique
for parameter optimization in VQE. Adopting a similar technique for
QAOA could improve parameter optimization and overall results.

The NchooseK approach offers a formulation that is both intuitive and
straightforward; however, it is impeded by degenerate minimum energy
solutions as a result of its dependence on soft constraint
penalties. In the future, we intend to improve the NchooseK algorithm
by dynamically modifying parameters, including soft constraint
penalties, D-Wave shot counts and annealing times to attain more
optimal outcomes.

Due to limited computational resources, the experiment was conducted
using only 4, 5, and 6 currency pairs. However, there is potential to
extend the algorithm to larger sets of currencies, which could uncover
more profitable currency arbitrage opportunities.

\section{Conclusion}

We effectively implemented quantum computing methodologies to address
the currency arbitrage problem employing both D-Wave quantum annealers
and gate-based quantum computers in conjunction with QAOA. The results
of our experiments indicate that D-Wave annealers exhibited a rapid
convergence to optimal solutions. In contrast, the implementation of
QAOA utilizing the most recent iteration of Qiskit demonstrated a
tendency to become trapped in local minima, leading to trading paths
that were suboptimal. Furthermore, it was noted that QAOA encountered
difficulties in meeting constraints both under NchooseK and standard
versions, resulting in an insufficient examination of currency pairs
and, at times, yielding erroneous outcomes.

A significant challenge in this study was the effective execution of
the QUBO model on both D-Wave quantum annealers and gate-based quantum
computers, such as IBM’s quantum processors. Implementing the QUBO on
D-Wave's hardware required the application of optimal embedding
techniques to effectively map our problem onto the machine's qubit
connectivity, thereby ensuring minimal error rates and maximizing
solution accuracy. Similarly, the implementation on gate-based quantum
computers necessitated the efficient decomposition of the QUBO into
circuits designed to minimize noise and gate errors. Successfully
navigating these challenges will facilitate a comparative analysis of
the outcomes from both platforms, thereby enabling an assessment of
the viability, scalability, and accuracy of each methodology.



\vspace{12pt}

\end{document}